\documentclass[seceq]{ptptex}

\usepackage{graphicx}
\usepackage{wrapft}
\usepackage{epsf}

\title{
Testing a model for the puzzling spin 0 mesons
}

\author{
Joseph \textsc{Schechter}%
}

\inst{
Physics Department, Syracuse University, Syracuse NY 
13244-1130 USA}

\abst{
 After a brief historical discussion of meson quantum numbers, we examine
 the possibility of additional internal meson structure. Experimental 
 tests of this structure using the semi-leptonic decays 
 of the $D_s^+$(1968) meson are discussed.
}

\begin{document}

\maketitle

\section{Further evolution of the Sakata Model}
The Sakata model \cite{rf:1} attempted to explain the ``zoo" of strongly interacting
``elementary particles" emerging in the 1950's by postulating 
that they were composites of the three low-lying  spin 1/2 fermions:

\begin{equation}
p,n,\Lambda.
\nonumber
\end{equation}

In this picture, mesons were considered to be objects
like $p{\bar n}$.This enables one to construct schematically 
all the observed hadrons.  It also suggests a natural
 SU(3) ``flavor" symmetry, first studied \cite{rf:2} at
 Nagoya.

A few years later the three fundamental hadronic fields were 
replaced by the fractionally charged quarks:
\begin{equation}
 u,d,s.
 \nonumber
 \end{equation}

Mesons are now considered to be objects like $u{\bar d}$. 

  It should be remarked that in addition to his profound insights into 
  elementary particle physics, Soichi Sakata championed a ``democratic"
  style of organization for physics research and education.
  
Moving forward, we note that three more quarks were found during
the great years for discovery between
1974 and 1995. This brings the total 
picture to: 

 \begin{equation}
 u,d,s,c,b,t.
 \nonumber
 \end{equation} 
  
  It is now easier to describe the quarks as:
  
  \begin{equation}
 q_a, \quad a=1\cdots 6
 \nonumber
 \end{equation}
  and raises  the question of whether any more will be found at LHC.

Of course, during this period it was also discovered that the strong dynamics 
is described by an ``SU(3) color" gauge theory so we must add a color index:

\begin{equation}
 q_{aA}, \quad a=1 \cdots 6, \quad A=1 \cdots 3 .
  \nonumber
 \end{equation}

 But we are still not done. If we regard this symbol as representing a 
 massless Fermi-Dirac field, we know that the left and right handed 
 projections enter differently into the unified electroweak theory. 
 Thus, we distinguish the two by agreeing to leave the left index alone and putting 
 a dot on the right index. 
 
  In this language, a spin zero meson made of a quark and an antiquark can 
  be schematically described as:  
 \begin{equation}
M_a^{\dot{b}} = {\left( q_{bA} \right)}^\dagger \gamma_4 \frac{1 +
\gamma_5}{2} q_ {aA}, \label{M}
\end{equation}
 
 Using a matrix notation, the decomposition in terms of scalar and
  pseudoscalar fields is:

\begin{equation}
M=S+i\phi, \quad M^{\dagger}=S-i\phi.
\label{m}
 \end{equation}

If all six quarks were massless, the symmetry of the color gauge 
theory Lagrangian would be:

\begin{equation}
SU(6)_L \times SU(6)_R  \times U(1)_{VECTOR}.
\nonumber
\end{equation}

Actually, the first three quarks are relatively light so the
reduced symmetry SU(3)$_L$ $\times$ SU(3)$_R$  $\times$ U(1)$_{V}$, 
while spontaneously broken, forms the basis of the chiral perturbation scheme 
which is successful at low energies.

\section{Considering the spin 0 mesons}

By counting, there are nine light (i.e. made as quark-antiquark composites
from the three lightest flavors) pseudoscalar mesons and nine light
scalar mesons. We learned at this conference that Sakata recommended
considering physics problems from two different perspectives. 

${\bf Theorist's\quad perspective:}$

 There are eight zero mass pseudoscalar Nambu-Goldstone bosons 
 and one heavier pseudoscalar boson (since the axial U(1) is 
 intrinsically broken by the axial anomaly and thus can't be spontaneouly 
broken).

  In order to make  chiral perturbation theory calculations,
  which automatically recover the ``current algebra' theorems as a 
  starting point, it is convenient to neglect the scalar mesons. This 
  is elegantly done by ``integrating them out".
  That has given rise to a belief that the nine scalar mesons should have
  infinite mass(so integrating them out would be rigorous) or at least should
  be very heavy.
  
${\bf Experimentalist's\quad perspective:}$

  Look at the data!
  
   This is not so easy. The starting point 
  is a partial wave analysis of pi-pi scattering in the I = J = 0
  channel. The real part, $R_0^0(\sqrt{s})$ is plotted both in Fig. 1a 
  and Fig. 1b, the experimental results corresponding to the points with error bars.
  Certainly the pattern is a complicated one.
  The fits shown by the solid lines 
 (M. Harada, F. Sannino, J.S.) \cite{hss96} include the particular contributions:
 
     a)chiral Lagrangian background
     
     b)rho meson background
     
     c)broad scalar (sigma) meson around 550MeV
     
     d)``f${_0}$(980)" candidate scalar meson (In Fig. 1a it is assumed that
     this meson decays completely into two pions while in Fig. 1b some decay
     into $K + {\bar K}$ is allowed.

A number of other groups \cite{other} have found similar results. 
  

\begin{figure}[t]
\begin{center}
\epsfysize = 5cm
\epsfbox{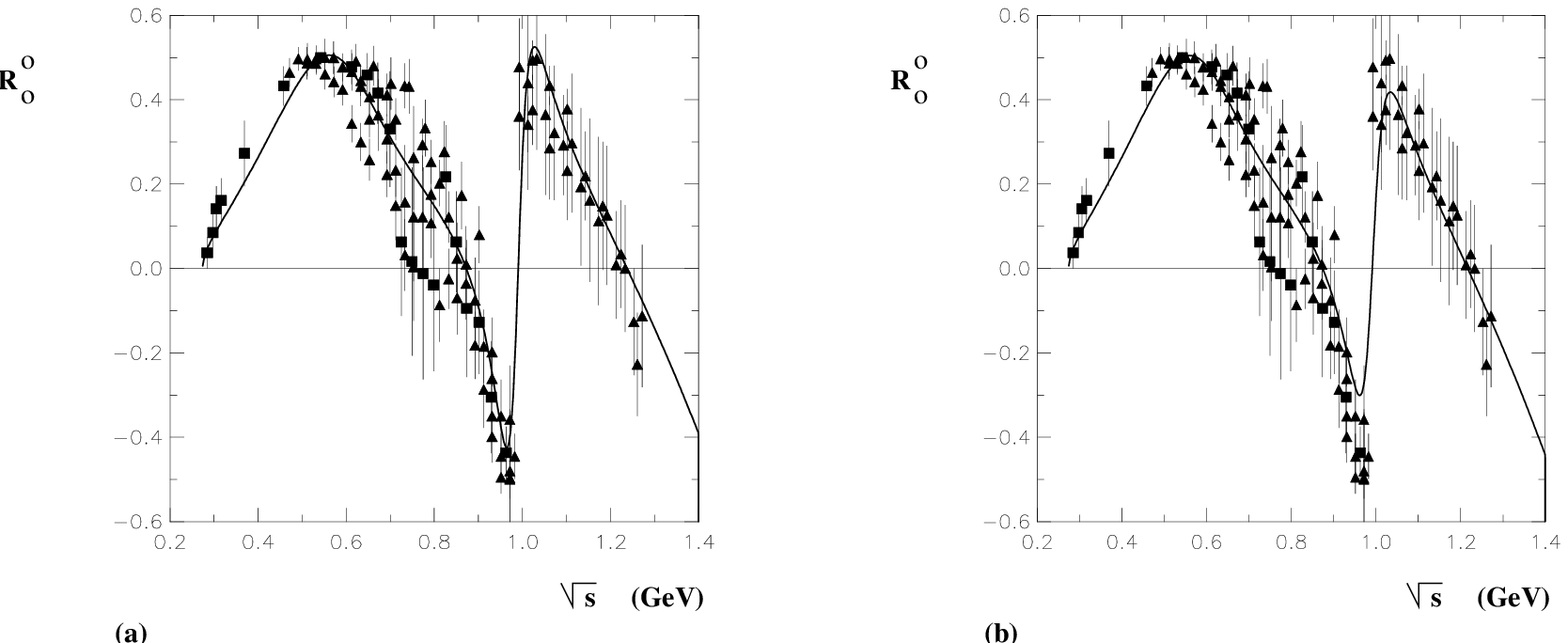}
\caption[]{
$R_0^0$
 }
\end{center}
\label{fig1}
\end{figure}


  For orientation, note that the relevant ``current algebra" theorem for this
  scattering gives the initial slope at threshold. It says nothing about the
  detailed pattern away from threshold. For fitting the experimental pattern
  away from threshold it turns out that  the rather light broad sigma state is
  crucial; without the sigma, the very large rho exchange contribution would
  force the amplitude quite soon above the unitarity bound, $R_0^0 < 1/2$.   
 
 Similar treatments of the pi-K scattering \cite{pik} and the pi-eta scattering \cite{pieta} have
  produced evidence for a spin 0 scalar strange meson multiplet (kappa) and 
  agreement with the experimental determination of another scalar resonance 
  with I = 1, the $a_0$(980). That finally yields a putative full nonet 
  of scalars:
  
  \begin{eqnarray}
  I&=&0:\quad m(\sigma)=550\hspace{.2cm} MeV
  \nonumber \\
  I&=&1/2:\quad m(\kappa)=800\hspace{.2cm} MeV
  \nonumber \\
  I&=&1:\quad m(a_0)=980\hspace{.2cm} MeV
  \nonumber \\
  I&=&0:\quad m(f_0)=980\hspace{.2cm} MeV 
  \nonumber \\
  \end{eqnarray}

This may be compared with the (most standard) vector meson nonet:   
  
  \begin{eqnarray}
  I&=&0:\quad m(\omega)=783\hspace{.2cm} MeV \hspace{.4cm}n{\bar n}
  \nonumber \\
  I&=&1:\quad m(\rho)= 776\hspace{.2cm} MeV\hspace{.4cm}n{\bar n}
  \nonumber \\
  I&=&1/2:\quad m(K^*)=980\hspace{.2cm} MeV\hspace{.4cm}n{\bar s}
  \nonumber \\
  I&=&0:\quad m(\phi)=1020\hspace{.2cm} MeV\hspace{.4cm}s{\bar s} 
  \nonumber \\
  \end{eqnarray}
    
 In this vector meson case, the usual quark-antiquark composition
 of each state is also displayed (s stands for a strange quark and
 n stands for either a u or a d quark). For the standard vector meson nonet, 
 the masses increase from the almost degenerate I = 0 and I = 1 particles 
 to the lone I = 0 particle $\phi$. Basically, this is due to the strange 
 quark being much heavier than the non-strange quarks u and d. However, for the 
 J = 0 scalar nonet candidates illustrated, the mass dependence is seen to be 
 exactly reversed!   
 A long time ago Jaffe \cite{jaffe} argued that a nonet made of two quarks 
 plus two antiquarks
 would have this reversed behavior (it simply reflects the SU(3) Clebsch
  Gordon coefficient $\epsilon_{abc}$ needed to couple two quarks to 
  an antiquark). We shall make use of this effect here.   
 
\section{Linear sigma models}
  
  In order to check the pi-pi, pi-K and pi-eta scattering results discussed
above, D. Black, A.H. Fariborz, S. Moussa, S. Nasri and JS \cite{BFMNS}recalculated them
in a relatively simple three flavor ${\it linear}$ sigma model using the field 
$M = S + i\phi$ mentioned above. 
 In that model the 
 unitarization 
of the partial wave amplitudes was accomplished using the K-matrix approach. This approach
has the ``advantage" that it does not introduce any additional parameters. It was found
that the same form of the complicated amplitude $R_0^0$ could be obtained. Similarly 
reasonable descriptions of 
the scattering partial waves involving the $\kappa$ and $a_0$ scalar states could also 
be obtained. But, by construction, ${\it both}$ the pseudoscalar and scalar nonets 
start out as quark-antiquark objects in the model.    

    To add confusion, there are some observed scalar and pseudoscalar 
    states which are not acommodated in this model. Also the lighter 
    scalar masses are much lower than where they are expected to be
    according to the reasonable non-relativistic quark model. In that model,
    the lowest mass nonets (below 1 GeV) are the pseudoscalars and the vectors. 
    The next highest nonets (somewhat above 1 GeV) are the scalars, tensors and
    two axial vectors (with different C properties). 
    
    So, the situation concerning the spin 0 chiral partners seems to
     call for clarification. For this purpose, 
    D. Black et al \cite{BFMNS} also proposed that there might be two 
    chiral spin 0 nonets - one of quark-antiquark type (M) and the other of   
    two quark- two antiquark type ($M^\prime = S^\prime + i \phi^{\prime})$) and 
    that they be allowed to mix with each other. This mixing is expected to lead 
    to level repulsion which could make the lighter scalars even lighter and 
    the heavier ones even heavier.  

       What would the schematic structure of the ``four quark" chiral nonet,
       $M^\prime$ look like? Assuming that $M^\prime$ 
       has the same chiral transformation property as $M$, there are three possibilities:

       ``Molecular" type:
  \begin{equation}
M_a^{(2) \dot{b}} = \epsilon_{acd} \epsilon^{\dot{b} \dot{e} \dot{f}}
 {\left( M^{\dagger} \right)}_{\dot{e}}^c {\left( M^{\dagger}
 \right)}_{\dot{f}}^d.
\label{Sandphi}
\end{equation}

Color triplet diquark - anti diquark type:
   \begin{equation}
   M^{(3){\dot f}}_g = (L^{gA})^\dagger R^{{\dot{f}}A},
   \nonumber
   \end{equation} 

where,
\begin{eqnarray}
L^{gE}&=&\epsilon^{gab}\epsilon^{EAB}q^T_{aA}C^{-1}\frac{1+\gamma_5}{2}q_{bB}
\nonumber \\
R^{{\dot g}E}&=&\epsilon^{{\dot g}{\dot a}{\dot b}}\epsilon^{EAB}q^T_{{\dot a}A}C^{-1}\frac{1-\gamma_5}{2}q_{{\dot b}B}
\nonumber
\end{eqnarray}

Color sextet diquark - anti diquark type:

 \begin{equation}
M_g^{(4) \dot{f}} = (L^g_{\mu\nu,AB})^\dagger R^f_{\mu\nu,AB}\quad,
\end{equation}

where,
\begin{eqnarray}
L^g_{\mu\nu,AB}&=&\epsilon^{gab}q^T_{aA}C^{-1}\sigma_{\mu\nu}\frac{1+\gamma_5}{2}q_{bB}
\nonumber \\
R^{\dot g}_{\mu\nu,AB}&=&\epsilon^{{\dot g}{\dot a}{\dot b}}q^T_{{\dot a}A}C^{-1}\sigma_{\mu\nu}\frac{1-\gamma_5}{2}q_{{\dot b}B}.
\nonumber
\end{eqnarray}

The distinction between molecular and diquark-antidiquark pieces is 
not fundamental because of the Fierz identity:
\begin{equation}
8M^{(2)\dot{b}}_a = 2M^{(3)\dot{b}}_a - M^{(4)\dot{b}}_a
\nonumber
\end{equation}

For the purpose of constructing a generalized linear sigma model containing both
quark- antiquark and diquark-antidiquark mesons we just need to assume that 
some unspecified linear combination of $M^{(2)}$,$M^{(3)}$ and $M^{(4)}$ is bound  
and may be designated as $M^{\prime}$. Note that $M$ and $M^{\prime}$ have different 
behaviors \cite{BFMNS} under the singlet axial transformations $U(1)_A$ 
even though they transform in the same way under $SU(3)_L\times SU(3)_R$.

\section{$M-M^\prime$ linear sigma model}

This complicated model has a number of aspects and was further
discussed in a series of papers \cite{FJS2005}-\cite{FJS2011C}.
See also\cite{seealso}. Similar perspectives 
are discussed in the papers\cite{diffshadings}.

We do not make any a priori assumptions about what are the quark-antiquark
and two quark-two antiquark contents of the 18 scalar and 18 pseudoscalar states
which emerge but let the model, with some experimental inputs, tell us the answer.

    The pieces of the ``toy" model Lagrangian include:
    
A)kinetic terms:

\begin{equation}
-\frac{1}{2}Tr(\partial_\mu M\partial_\mu M^\dagger) 
-\frac{1}{2}Tr(\partial_\mu M^\prime\partial_\mu M^{\prime\dagger)},    
\nonumber
\end{equation}    

B)symmetry breaking quark mass terms:

\begin{equation}
-2Tr(AS),
\nonumber
\end{equation}
where $A=diag(A_1,A_2,A_3)$ are proportional to the three light 
quark masses.

C)chiral invariant interaction terms:

\begin{equation}
-c_2Tr(MM^\dagger) +c_4^aTr(MM^{\dagger}MM^{\dagger}) +d_2Tr(M^{\prime}M^{\prime\dagger})
+c_3^a(\epsilon_{{\dot a}{\dot b}{\dot c}}\epsilon^{def}M_d^{\dot a}M_e^{\dot b}M_f^{\dot c}
+h.c.))
\nonumber
\end{equation}

(There are about 20 renormalizable terms \cite{FJS2005}; we just kept those with
 8 or less underlying quarks.)
 
D) terms to mock up the U(1) axial anomaly:  

\begin{equation}
c_3[\gamma_1ln(det M/det M^\dagger) 
+(1-\gamma_1)ln((Tr(MM^{\prime\dagger})/Tr(M^\prime M^\dagger))
\nonumber
\end{equation}

Terms of ${\it both}$ the $det M$ and $ Tr(MM^{\prime\dagger}) $ types 
appear \cite{3fl} in the 3-flavor 't Hooft - type 
instanton calculation.

E)The needed vacuum values (assuming isospin invariance) are:

\begin{equation}
<S_1^1> = <S_2^2>, <S_3^3>, <S^{\prime 1}_1>,<S^{\prime 3}>
\nonumber
\end{equation}
 
F)The following 8 inputs are used:

\begin{eqnarray}
    F_\pi &=&<S_1^1> + <S_2^2> =131 MeV, \quad A_3/A_1 = 20- 30,
    \nonumber \\
  m[a_0(980)] &=& 984.7 \pm 1.2 MeV,\quad  m[a_0(1450)]=1474 \pm 19 MeV,
  \nonumber \\
  m_\pi &=&137 MeV,\quad m[\pi(1300)] =1300 \pm 100 MeV,
  \nonumber 
\end{eqnarray}
and in addition:    two mass parameters for the four I=0 $\eta's$.
  
\section{Predicted meson properties in the $M-M^\prime$ model }

    In the $M -M^\prime$ model there are eight different 
   pseudoscalar isomultiplets. Their tree level masses  are
   displayed in the table below and are clearly a mixture of the input
   masses and a few predictions. However, all eight of the 
   ``two quark" vs. ``four quark" percentages are predicted by the model. 
   Not surprisingly, the lower mass particles of each isospin turn 
   out to be dominantly of quark- antiquark type. Note that all four 
   of the I = 0 particles (the $\eta$'s )mix with each other to some extent.
   Of course there are enough pseudoscalars to fill two nonets.

\begin{table}[htbp]
\begin{center}
\begin{tabular}{c|c|c|c}
\hline \hline
State & \,\, ${\bar q} q$\% \, \, & \, ${\bar q} 
{\bar q} q q$\%\, & \, $m$ (GeV)
\\
\hline
\hline
$\pi$      & 85  & 15 & 0.137\\
\hline
$\pi'$      & 15  & 85 & 1.215\\
\hline
$K$          & 86  & 14 & 0.515\\
\hline
$K'$          & 14  & 86 & 1.195\\
\hline
$\eta_1$   &  89   &  11   & 0.553 \\
\hline
$\eta_2$   &  78   &  22   & 0.982\\
\hline
$\eta_3$   &  32   &  68   &  1.225 \\
\hline
$\eta_4$   &  1   & 99     & 1.794\\
\hline
\hline
\end{tabular}
\end{center}
\caption[]{
Typical predicted properties of pseudoscalar
states:
${\bar q} q$ percentage (2nd
column), ${\bar q} {\bar q} q q$ (3rd
column) and masses (last column).}
\label{phi_content_1215}
\end{table}

The properties of the scalar mesons in the $M -M^\prime$ model are displayed
in the table below. In this case the only mass inputs were
for the $a$ and $a^\prime$ isotriplets; all the other masses and the 
two quark vs. four quark percentages are predictions of the model.
Here the situation is opposite to that of the pseudoscalars. The lower 
lying states are predominantly two quark - two antiquark ( or``four quark")
type. For example, 
the lighter isovector, $a$ is 76 per cent ``four quark" while the heavier 
isovector, $a^\prime$ is 24 per cent ``four quark".

  The famous $\sigma$ = $\sigma_1$ is 40 percent quark- antiquark and 
  60 per cent ``four quark". The $f_0$(980) is 95 percent of 
  ``four quark" type. 

 Note that these masses are ``tree level" ones. For the $\sigma$, the 
 unitarity corrections (which involve computing the $\pi\pi$ scattering 
 amplitude in the model) reduce \cite{FJSS2011B} the predicted mass 
 to 477 MeV.

\begin{table}[htbp]
\begin{center}
\begin{tabular}{c|c|c|c}
\hline \hline
State &\,\, ${\bar q} q$\%\,\,& \, ${\bar q} 
{\bar q} q q$\% \, & \, $m$ (GeV)
\\
\hline
\hline
$a$        &  24 &  76 & 0.984   \\
\hline
$a'$        &  76 &  24 & 1.474   \\
\hline
$\kappa$  &   8  &  92  & 1.067  \\
\hline
$\kappa'$  &   92  &  8  & 1.624 \\
\hline
$\sigma_1$     &  40    &  60  & 0.742  \\
\hline
$\sigma_2$     &  5   &    95  & 1.085 \\
\hline
$\sigma_3$     &  63   &   37  & 1.493 \\
\hline
$\sigma_4$     &  93   &   7   & 1.783 \\
\hline
\hline
\end{tabular}
\end{center}
\caption[]{ Typical predicted properties of scalar 
states: 
${\bar q} q$ percentage (2nd 
column), ${\bar q} {\bar q} q q$ (3rd 
column) and masses (last column).
 }
\label{s_content_1215}
\end{table}

\section{Experimental information on scalar mesons}
Typically, it comes from partial wave analyses of
 scattering processes. Another source arises from Dalitz analyses 
 of multiparticle final states in non-leptonic weak decays.
 
 Recently, the CLEO Collaboration obtained \cite{cleo} a simple 
 neat determination of the mass and the width of the $f_0(980)$ 
 [$\sigma_2$ in the notation above] from the semi-leptonic decay
of a charmed meson:

\begin{equation}
D_s^+(1968) \rightarrow f_0(980) + e^+ + \nu_e.
\nonumber
\end{equation}

This process correponds to the ``quark" picture in Fig. 2.

\begin{figure}[htbp]
\centering
\rotatebox{0}
{\includegraphics[width=7cm,clip=true]{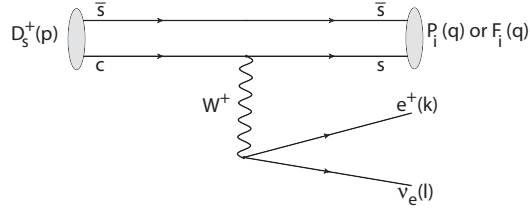}}
\caption[]{
$D_s$ decay.
}
\label{Ds}
\end{figure}

\section{Predicting some semi-leptonic decay widths}

   It seems interesting that in addition to the decay into what 
   we called $\sigma_2$, there should also exist decays into $\sigma_1$,
    $\sigma_3$ and $\sigma_4$
   which are easily predictable in the model. All four of these 
   particles arise as mixtures of the standard basis states 

\begin{eqnarray}
f_a&=&\frac{S^1_1+S^2_2}{\sqrt{2}} \hskip .7cm
n{\bar n},
\nonumber  \\
f_b&=&S^3_3 \hskip .7cm s{\bar s},
\nonumber    \\
f_c&=&  \frac{S'^1_1+S'^2_2}{\sqrt{2}}
\hskip .7cm ns{\bar n}{\bar s},
\nonumber   \\
f_d&=& S'^3_3
\hskip .7cm nn{\bar n}{\bar n}.
\label{fourbasis}
\end{eqnarray}

We denote $f$ as the four component vector: $(f_1,f_2,f_3,f_4)$.
 For typical
values of the model's input parameters 
 the mass eigenstates, $\sigma_i$
 make up a four vector, $\sigma$ = $L_0^{-1}f$ with,

\begin{equation}
(L_o^{-1})  =
\left[
\begin{array}{cccc}
0.601  &  0.199 &  0.600  &    0.489 \\
-0.107   &  0.189 &  0.643  &     -0.735  \\
0.790   &  -0.050&  -0.391  &   -0.470  \\
0.062 &  -0.960 &   0.272  &   -0.019\\
\end{array}
\right]
\label{4by4rots}
\end{equation}

The physical states are identified, with nominal mass values, as
\begin{equation}
\sigma  =
\left[
\begin{array}{c}
f_0(600)   \\
f_0(980)  \\
f_0(1370)  \\
f_0(1800)\\
\end{array}
\right]
\label{fphys}
\end{equation}

Similarly, there are four predictions of the model for decays 
of the $D_s^+$ into the pseudoscalar singlets $\eta_i$ + leptons.

     The required hadronic information consists of the vector (for 
     decays into the $\eta_i$ and the axial vector (for decays into the
     $\sigma_i$) Noether currents of the model. For the three flavor model
      these currents take the well known forms: 

 \begin{eqnarray}
      V_{\mu a}^b(total)&=&V_{\mu a}^{b}+V^{\prime b}_{\mu a}, \nonumber  \\
       A_{\mu a}^b(total)&=&A_{\mu a}^{b}+A_{\mu a}^{\prime b}.
       \label{totalsu3currents}
       \end{eqnarray}
       
 \begin{eqnarray}
    V_{\mu a}^b&=&i\phi_a^c\stackrel{\leftrightarrow}{\partial_\mu}\phi_c^b +
    i\tilde{S}_a^c\stackrel{\leftrightarrow}{\partial_\mu}\tilde{S}_c^b
    +i(\alpha_a-\alpha_b)\partial_\mu\tilde{S}_a^b,
\nonumber \\
A_{\mu a}^b&=&\tilde{S}_a^c\stackrel{\leftrightarrow}{\partial_\mu}\phi_c^b -
    \phi_a^c\stackrel{\leftrightarrow}{\partial_\mu}\tilde{S}_c^b
    +(\alpha_a+\alpha_b)\partial_\mu\phi
    _a^b,
\label{su3currents}
\end{eqnarray}      
       
          \begin{eqnarray}
    V_{\mu a}^{\prime b}&=&i\phi_a^{\prime c}\stackrel{\leftrightarrow}{\partial_\mu}\phi_c^{\prime b} +
    i\tilde{S}_a^{\prime c}\stackrel{\leftrightarrow}{\partial_\mu}\tilde{S}_c^{\prime b}
    +i(\beta_a-\beta_b)\partial_\mu\tilde{S}_a^{\prime b},
\nonumber \\
A_{\mu a}^{\prime b}&=&S_a^{\prime c}\stackrel{\leftrightarrow}{\partial_\mu}\phi_c^{\prime b} -
    \phi_a^{\prime c}\stackrel{\leftrightarrow}{\partial_\mu}\tilde{S}_c^{\prime b}
    +(\beta_a+\beta_b)\partial_\mu\phi
    _a^{\prime b},
\label{primesu3currents}
\end{eqnarray}  
       
  Note that $\alpha_a = <S_a^a>$ and $\beta_a = <S_a^{{\prime}a}>$. 
  Also ${\tilde S}_a^b = S_a^b - < S_a^b>$.    
       
\section{Noether currents in the four flavor case}       
       
       Clearly, an extension is needed to accomodate the charmed $D_s^+$ particle 
       required to calculate the desired semi-leptonic decay rates. Also,
        it at first seems dubious to consider the fourth heavy quark in 
        the same Lagrangian as the three light quarks. However for the present 
        calculation, only the Noether currents are needed and these depend only on the 
       kinetic terms which are independent of mass related parameters. 
       
       So, one's first thought is to just sum 1 - 4 instead of 1 - 3 in the currents above.
       However there is still a problem with this simple extension.
        A fourth quark flavor will not allow
       the construction of a two quark-two antiquark state which has the same chiral SU(4)
       transformation property as the one quark - one antiquark state with
        which it is supposed to mix. For example,trying a ``molecule" form would result in
        
        \begin{equation}
        M_{ag}^{(2){\dot b}{\dot h}} = \epsilon_{agcd}\epsilon^{{\dot b}{\dot h}{\dot c}{\dot f}}
        (M^\dagger)_{\dot e}^c(M^\dagger)_{\dot f}^d.
        \nonumber
        \end{equation}
        
        But, instead of transforming under $SU(4)_L \times SU(4)_R$ as $(4,{\bar 4})$  it transforms as 
        $(6,{\bar 6})$ owing to the two sets of antisymmetric indices which appear.
         It may be shown \cite{{FJSS2011A}} that 3 flavors are special in allowing the 
         kind of mixing which preserves the underlying chiral symmetry.
            
       Thus, we assume that there are no two quark - two antiquark 
       components for the mesons containing a charm quark. The kinetic 
       terms for the model may then be written as:

 \begin{equation}
    {\cal L} = -\frac{1}{2} Tr^4(\partial_\mu M \partial_\mu M^\dagger)
               -\frac{1}{2} Tr^3(\partial_\mu M^\prime \partial_\mu M^{\prime\dagger}),
      \label{hybridlag}
       \end{equation}

      where the meaning of the superscript on the trace symbol is
      that the first term should be summed over the heavy quark index as
      well as the three light indices. This stands in contrast to the second term
      which is just summed over the three light quark indices pertaining to the
      two quark - two antiquark field $M^\prime$. Since the Noether
      currents are sensitive only to these
      kinetic terms in the model, the vector and axial vector currents with
      flavor indices 1 through 3 in this model are just the same as in
       Eq.(\ref{totalsu3currents}) above. However if either or both flavor
       indices take on the value 4 (referring to the heavy flavor) the current
       will only have contributions from the field $M$. This should be clarified
       by the following example,

\begin{eqnarray}
 V_{\mu 4}^a(total)&=&V_{\mu 4}^a=i\phi_4^c\stackrel{\leftrightarrow}{\partial_\mu}\phi_c^a +
    i S_4^c\stackrel{\leftrightarrow}{\partial_\mu}S_c^a,
    \nonumber \\
 A_{\mu 4}^a(total)&=&A_{\mu 4}^a=S_4^c\stackrel{\leftrightarrow}{\partial_\mu}\phi_c^a -
    \phi_4^c\stackrel{\leftrightarrow}{\partial_\mu}S_c^a.
\label{heavycurrents}
\end{eqnarray}
    Here the unspecified
    indices can run from 1 to 4.
    
   Note that the currents do not contain any unknowns; their normalization is 
   given by the component which is the electric current (i.e. ``conserved vector 
   current" hypothesis). Then the unintegrated decay widths into any of the four isoscalar
   $0^+$ mesons or four isoscalar $0^-$ mesons is given by,
   
   \begin{equation}
\frac{d\Gamma_i}{d|\bf{q}|}= \frac{G_F^2|V_{cs}|^2}{12\pi^3} \left\{ \begin{array}{c}
((R_0)_{2i})^2\\
((L_0)_{2i})^2
\end{array}\right\} m(D_s)\frac{|{\bf q}|^4}{q_0}.
\label{udw}
\end{equation}

    where $q_\mu$ is the final meson four momentum and $V_{cs}$ is the 
   Kobayashi Maskawa matrix element. $R_0$ is the pseudoscalar analog of 
   $L_0$ introduced in Eq. (7.2) for the scalars. 
   
Table III summarizes the calculations of the predicted widths, for $D_s^+$
decays into the four pseudoscalar singlet mesons ($\eta_1=\eta$(547), $\eta_2=\eta$(982),
$\eta_3=\eta$(1225), $\eta_4=\eta$(1794). Notice that the listed masses, $m_i$ are the ``predicted" ones
in the present model.

\begin{table}[htbp]
\begin{center}
\begin{tabular}{c||c|c||c}
\hline \hline
$m_i$ (MeV) & $(R_0)_{2i}$ & $(q_{max})_i$ (MeV) & $\Gamma_i$ (MeV)
\\ \hline
553 & 0.661 & 906.20 & 4.14 $\times$ 10$^{-11}$
\\ \hline
982 & 0.512 & 739.00 & 7.16 $\times$ 10$^{-12}$
\\ \hline
1225 & -0.546 & 602.74 & 2.57 $\times$ 10$^{-12}$
\\ \hline
1794 & 0.051 & 166.31 & 2.65 $\times$ 10$^{-17}$
\\ \hline
\hline
\end{tabular}
\end{center}
\caption[]{pseudoscalars.}
\label{pscalars}
\end{table}
    
Table IV, with the same conventions, summarizes the calculations of the predicted widths for $D_s^+$
decays into the four scalar singlet mesons [${(\sigma_1,\sigma_2,\cdots)=(\sigma,f_0(980),\cdots)}$]
 and leptons.

\begin{table}[htbp]
\begin{center}
\begin{tabular}{c||c|c||c}
\hline \hline
$m_i$ (MeV) & $(L_0)_{2i}$ & $(q_{max})_i$ (MeV) & $\Gamma_i$ (MeV)
\\ \hline
477 & 0.199 & 933.23 & 4.56 $\times$ 10$^{-12}$
\\ \hline
1037 & 0.189 & 710.79 & 7.80 $\times$ 10$^{-13}$
\\ \hline
1127 & -0.050 & 661.30 & 3.62 $\times$ 10$^{-14}$
\\ \hline
1735 & -0.960 & 219.21 & 3.85 $\times$ 10$^{-14}$
\\ \hline
\hline
\end{tabular}
\end{center}
\caption[]{scalars.}
\label{scalars}
\end{table}

Experimental data exist for only three of these eight decay modes:

\begin{eqnarray}
\Gamma (D_s^+\rightarrow \eta e^+ \nu_e)&=&(3.5 \pm 0.6) \times 10^{-11}\quad \rm{MeV}\nonumber\\
\Gamma (D_s^+\rightarrow \eta^\prime e^+ \nu_e)&=&(1.29 \pm 0.30) \times 10^{-11}\quad \rm{MeV}\nonumber\\
\Gamma (D_s^+\rightarrow f_0 e^+ \nu_e)&=&(2.6 \pm 0.4) \times 10^{-12}\quad \rm{MeV}
\label{expw}
\end{eqnarray}    
    
  It is encouraging that even though the calculation utilized the simplest model
    for the current and no arbitrary parameters were introduced, the prediction for the
    lightest hadronic mode,
    $\Gamma (D_s^+\rightarrow \eta e^+ \nu_e)$ agrees with the measured value. In the case of
    the decay $D_s^+\rightarrow \eta e^+ \nu_e$ the predicted width is about 30$\%$ less than the
    measured value. For the mode $D_s^+\rightarrow f_0(980) e^+ \nu_e$ the predicted value is about
    one third the measured value. Conceivably, considering the large predicted width into the very broad
    sigma state centered at 477 MeV, some of the higher mass sigma events might have been counted as $f_0$(980)
     events, which would improve the agreement. It would be very interesting to obtain
     experimental information about
     the energy regions relevant to the other five predicted isosinglet modes.

     Furthermore, varying the
     particular choice for the quark mass ratio $A_3/A_1$ and the precise mass of the very broad $\Pi$(1300)
     resonance within the allowable ranges can lead to a satisfactory fit to experiment, as shown 
     in \cite{FJS2011C}.
      
      \section{Summary}
      
         The light spin 0 pseudoscalar mesons appear to be 
         of $q{\bar q}$ type.
         
         The light spin 0 scalar mesons appear to be of $qq{\bar q}{\bar q}$ type.
         
         Chiral symmetry is a symmetry of massless QCD so these mesons should
         be reasonably approximated as chiral partners. Then how can we reconcile 
         their different compositions?
         
          Proposed solution: Introduce a chiral $q{\bar q}$ multiplet ${\it and}$ a 
         chiral $qq{\bar q}{\bar q}$ multiplet. They mix and the lightest pseudoscalars
         are mainly $q{\bar q}$ while the lightest scalars are mainly $qq{\bar q}{\bar q}$.
         
         Semi-leptonic decays of the heavy mesons seem to provide useful 
          experimental information for checking this picture.

\section{Acknowledgements} I am pleased to acknowledge the important roles of my
collaborators Deirdre Black, Amir Fariborz, Renata Jora and Naeem Shahid in developing this 
model.
I would like to thank the conference organizers for their gracious hospitality in
Nagoya. This work was supported in part by the U.S. DOE under Contract No. DE-FG-02-85ER40231.

%

\end{document}